\documentclass[noshowpacs,pre,preprintnumbers,superscriptaddress,amsmath,amssymb,floatfix]{revtex4}

\usepackage{amsmath}
\usepackage[caption=false]{subfig}
\usepackage{graphicx}
\usepackage{color,soul}
\usepackage{placeins}
\usepackage{tikz}
\usepackage{mathtools}
\usepackage{amsthm}
\usepackage{xkeyval,xcolor}
\usepackage{sidecap}

\newcommand{\av}[1]{\langle #1 \rangle}
\newcommand{\dd}{\mathrm{d}}

\newcommand{\nn}{\mathcal{N}}

\begin{document}
\title{Impact of food distribution on lifetime of a forager with or without sense of smell}

\author{Hillel Sanhedrai}
\affiliation{Department of Physics, Bar-Ilan University, Ramat Gan, Israel}
\author{Yafit Maayan}
\affiliation{Department of Mathematics, Jerusalem College of Technology (JCT), Jerusalem, Israel}

\date{\today}

\begin{abstract}
	Modeling foraging via basic models is a problem that has been recently investigated from several points of view. However, understanding the effect of the spatial distribution of food on the lifetime of a forager has not been achieved yet.
	We explore here how the distribution of food in space affects the forager's lifetime in several different scenarios. We analyze a random forager and a smelling forager in both one and two dimensions. We first consider a general food distribution, and then analyze in detail specific distributions including constant distance between food, certain probability of existence of food at each site, and power-law distribution of distances between food. For a forager in one dimension without smell we find analytically the lifetime, and for a forager with smell we find the condition for immortality. In two dimensions we find based on analytical considerations that the lifetime ($T$) scales with the starving time ($S$) and food density ($f$) as $T\sim S^4f^{3/2}$.
	
\end{abstract}

\maketitle

\section{Introduction}

Optimization of foraging for food spread in space 
is a problem that has been widely studied
\cite{stephens1986foraging,pyke1984optimal,benichou2011intermittent,mueller2011integrating}. Many studies claim that stochastic search yields optimal results  \cite{oaten1977optimal,green1984stopping,hein2012sensing} and that random walks or L\'evy flights can be used to model the forager behavior \cite{viswanathan1996levy,viswanathan1999optimizing,benichou2005optimal,lomholt2008levy,edwards2007revisiting}.  Several models for a forager's movement behavior have been proposed including some that are based on stimuli, memory, and cues from fellow foragers \cite{mueller2011integrating,bracis2015memory,vergassola2007infotaxis,martinez2013optimizing}.

Recent work has suggested a new model where a forager performs a random walk, however the food is explicitly consumed until the forager starves to death \cite{benichou2014depletion,benichou2016role}. In this model, the forager begins at some point on a lattice where each site contains a unit of food. The forager then moves and eats the food at the discovered site, leaving no remaining food in this site. It continues to move throughout the region either returning to sites without food or eating food at newly visited sites. If the forager walks $S$ steps without finding food and eating, it starves to death. Notably, this process leads to inherent desertification \cite{reynolds2007global,weissmann2014stochastic}, as the forager eventually creates a desert of visited sites among which it could move until starvation. Later work  expanded this model to cases where the food renews after some time \cite{chupeau2016universality}, where the forager eats only if it is near starvation \cite{rager2018advantage,benichou2018optimally}, and where the forager walks preferentially in the direction of a nearby site with food \cite{bhat2017does,bhat2017starvation}.

Another recent study \cite{sanhedrai2019epl} has extended the starving forager models to a forager with an explicit sense of \emph{smell} that extends to potentially longer ranges \cite{fagan2017perceptual}. The contribution of an individual food site to the overall smell in a given direction decays with its distance $d$ from the forager. While actual patterns of odor diffusion are turbulent and vary in time in highly complex ways \cite{celani2014odor}, it can assumed to be simplified by considering two realistic cases: power-law decay with distance and exponential decay of smell. 
The likelihood of the forager to walk in each direction is proportional to the total smell in that direction.

A recent study \cite{benichou2014depletion} has shown that the lifetime of a forager, $T$, in 1D scales linearly with its starving time, $S$, the time it can live without food. In 2D, however, it scales approximately as $T\sim S^2$. Another study \cite{sanhedrai2019epl} has shown recently that when there is a long range smell in 1D, then under some conditions the probability to live forever exists. These studies considered the case where initially all the space is full of food. Here we consider several cases of food distributions in space. We examine how different more realistic distributions impact the life time of the forager. We find for a random forager a general scaling function that includes the density of food. Moreover, we also find that for a smelling forager in 1D, the chance of immortality highly depends on food distribution. In addition, for a forager walking in two dimensions we find that the existence of long range smell increases the forager's lifetime dramatically from $T \sim S^2$ to $T \sim S^4$.


\section{One dimension - random forager} \label{sec: random 1d}
\begin{figure}[h]
	\centering
	\includegraphics[width=0.7\linewidth]{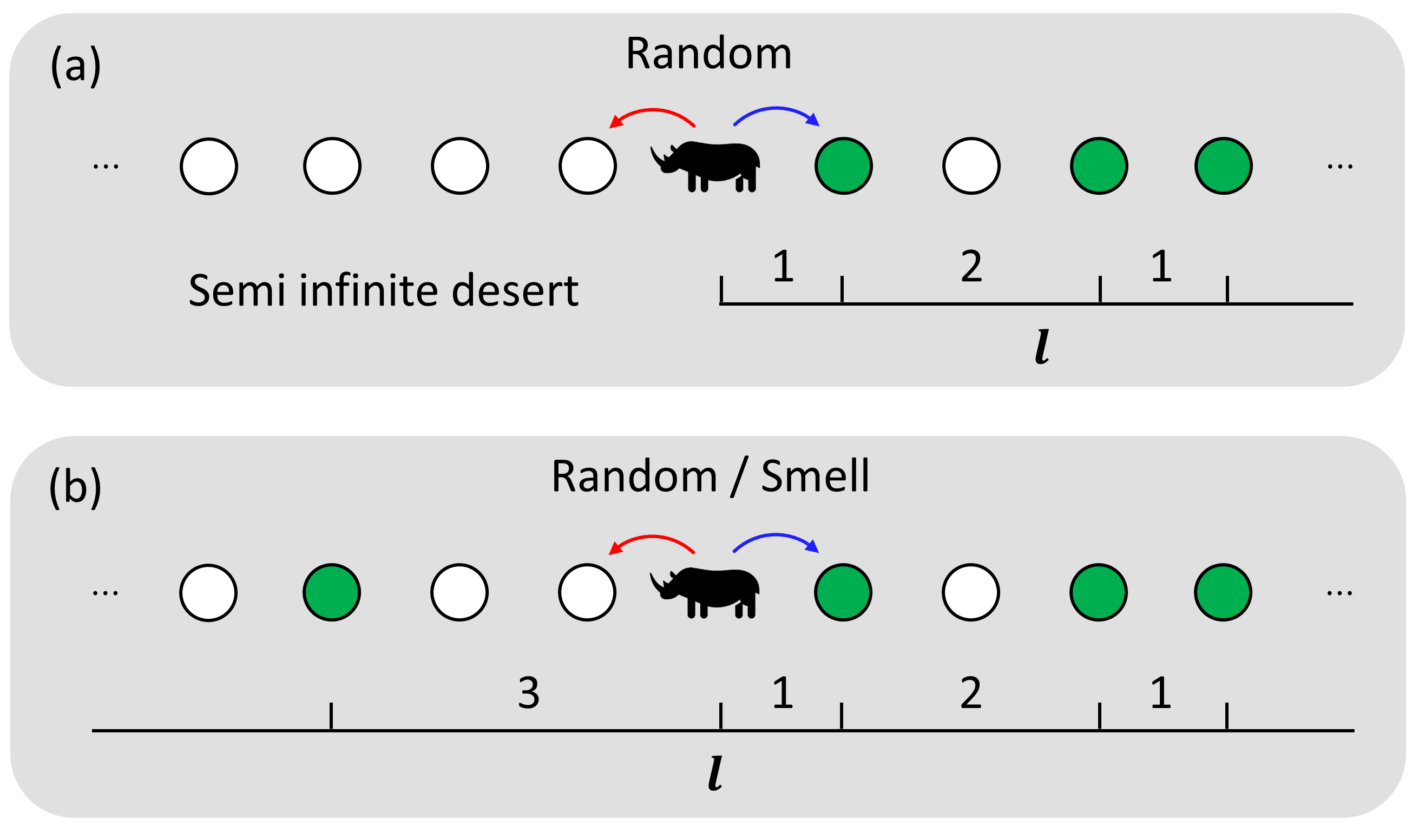}
	\caption{{\bf Illustration of forager in one dimension} (a) Here all food is in the right of forager (filled circles), whereas in the left there is a semi infinite desert. Between food there are distances, $l$, distributed according to $P(l)$. The forager walks randomly. This case is studied in Sec. \ref{sec: random 1d}. (b) In this scenario food is located in both directions. Here we study in detail the case of forager with long range smell in Sec. \ref{sec: smell 1d}.
	}
	\label{fig: model illustration}
\end{figure}

We aim in this chapter to find out how the distribution of food in space influences the lifetime of forager in one dimension, see illustration in Fig. \ref{fig: model illustration}.
First, we consider a forager walking \emph{randomly} on a one dimensional lattice. For getting full analytic solution we analyze the case of a semi-infinite desert. In this scenario, at the beginning there is food only at one side of the forager, while the other side is a semi-infinite desert.
We assume that between positions of food there is a distance $l$ distributed according to an arbitrary distance distribution $P(l)$, and the forager walks randomly. If the forager makes $S$ steps without reaching any food it starves and dies.

We are interested in the following quantities, the mean life time of the forager, $T$, and $\nn$, the expectation value of number of meals the forager consumed during its lifetime. We study also $\tau$, the expected time between meals given the next meal occurs. To this end, we should first evaluate $F(t)$, the likelihood of the forager to get food for the first time at step $t$ after it ate. It is well known \cite{feller1968book, redner2001book_first} that the generating function of $F(t,l)$, the first passage time probability to be at $x=0$ starting at $x=l$, is
\begin{equation}
\mathcal{F}(z,l) = \alpha(z)^l,
\end{equation}   
where
\begin{equation}
\alpha (z) = \frac{1 - \sqrt{1-z^2}}{z}.
\end{equation}
Then, we consider $x=0$ as the site with the closest food and $x=l$ as the site where the last meal happened.

Next, we note that the probability that the closest food is at distance $l$ given that the forager just ate is $P(l)$. \\
Thus, the first passage probability, $F(t)$ is 
\begin{equation}
F(t) = \sum_{l=1}^{\infty}  P(l) F(t,l),
\end{equation}
resulting in,
\begin{equation} \label{eq: F}
\mathcal{F}(z) = \sum_{l=1}^{\infty} P(l) \mathcal{F}(z,l)  
= \sum_{l=1}^{\infty} P(l) \alpha(z)^l  
= G(\alpha(z)),
\end{equation}
where 
\begin{equation}
G(x)=\sum_{l=1}^{\infty}P(l)x^l,
\end{equation}
is the generating function of the distance distribution $P(l)$. \\
Note that if the distance $l$ is always one, then $G(x)=x$, and $\mathcal{F}(z)=\alpha(z)$, which converges to the well known result for the scenario where space is filled with food \cite{bhat2017starvation}. \\

After having $\mathcal{F}(z)$, following the steps in \cite{bhat2017starvation} (see also Appendix \ref{sec: app T,tau,N by F}) we obtain for the average number of meals, $\nn$, and for the average time between meals, $\tau$,
\begin{align} \label{eq: N}
& \nn = \frac{{E(S)}}{1-{E(S)}},
\\
& \tau = \frac{\pi(S)}{E(S)}.
\label{eq: tau}
\end{align}
Thus, the average lifetime is
\begin{gather} \label{eq: T}
T  = \tau \nn +S = \frac{\pi(S)}{1-E(S)} + S,
\end{gather}
where $E(S)$ is derived from the generating function $\mathcal{E}(z) = \mathcal{F}(z)/(1-z)$, and the generating function of $\pi(S)$ is $\Pi(z) = z  \mathcal{F}'(z)/(1-z)$, where $\mathcal{F}(z)$ is given in Eq. \eqref{eq: F}.

To conclude, given the distribution of food in space $P(l)$, we find the lifetime, $T$, and the number of meals, $\nn  $. The term which depends directly on $P(l)$ and determines $T$ and $\nn  $ is $G(x)$. In the next Secs. we discuss three specific cases of food distribution having three different functions for $G(x)$.

\subsection{Asymptotic behavior for large $S$}
For finding the behavior of $\nn, \tau, T$ for large $S$ we analyze the asymptotic behavior of Eqs. (\ref{eq: N}), (\ref{eq: tau}) and (\ref{eq: T}) by expanding the corresponding generating functions in the limit $z\to1$ and using the Tauberian Theorems \cite{feller-vol-2}. For a more detailed analysis see Appendix \ref{sec: app large S}. We show that for the leading term, the only property of food distribution which matters is the mean distance between food, $\av{l}$, in case it is finite. We denote the density of food by $f=1/\av{l}$. We wish to get $\mathcal{F}(z)$, Eq.~\eqref{eq: F}, which determines all quantities. Because $\mathcal{F}(z) = G(\alpha(z))$ we treat first $\alpha(z)$, and then $G(x)$.
An expansion of $\alpha$ where $z\to1$ gives, $\alpha(z) \sim 1- \sqrt{2}\sqrt{1-z}$. Therefore, we analyze $G(x)$ for $x\to 1$. $G(1)=1$ due to normalization, and if the mean distance, $\av{l}$, is finite, $G'(1)=\av{l}$, and then using Taylor expansion, 
\begin{gather} \label{eq: G asimp finite}
	G(x) \sim  1 - \av{l}(1-x).
\end{gather}
Having $G(x)$ we obtain
\begin{align*}
\mathcal{F}(z) = G(\alpha(z)) \sim 1-\av{l}(1-\alpha(z))
\sim 1-\av{l}\sqrt{2}\sqrt{1-z} .
\end{align*}
Using this, we can derive all other quantities (see Appendix \ref{sec: app large S}) and obtain,
\begin{equation} \label{eq: N,tau,T asimp}
\begin{aligned}
& \nn   \sim   \sqrt{\frac{\pi}{2}} \frac{1}{\av{l}} \sqrt{S} \sim \sqrt{\frac{\pi}{2}} f \sqrt{S} ,
\\
& \tau \sim \sqrt{\frac{2}{\pi}} \av{l} \sqrt{S} \sim \sqrt{\frac{2}{\pi}} \frac{1}{f} \sqrt{S} ,
\\
& T = \nn \tau + S \sim  2S.
\end{aligned}
\end{equation}
However, if $\av{l}$ is infinite, and we assume that the distance distribution behaves according to $P(l)\sim l^{-(1+\beta)}$, where $0<\beta<1$ such that $\av{l}=\infty$, then unlike Eq. \eqref{eq: G asimp finite},
\begin{gather}
G(x) \sim 1 - A(1-x)^{\beta}.
\end{gather}
This result leads to
\begin{align*}
	\mathcal{F}(z) = G(\alpha(z)) \sim 1-A(1-\alpha(z))^{\beta}
	\sim 1-2^{\beta/2}A(1-z)^{\beta/2},
\end{align*}
what provides,
\begin{equation} \label{eq: N,tau,T beta asimp}
\begin{aligned}
& \nn \sim   \frac{\Gamma(\beta/2)}{2^{\beta/2}A} S^{\beta/2} ,
\\
& \tau \sim \frac{\beta 2^{\beta/2}A}{2\Gamma(2-\beta/2)}S^{1-\beta/2} ,
\\
& T \sim \left(\frac{\Gamma(1+\beta/2)}{\Gamma(2-\beta/2)}+1\right)S .
\end{aligned}
\end{equation}
Next we consider an edge case where $\beta=1$, which presents an infinite average distance between food as well. In this case we get a logarithmic correction, as follows,
\begin{equation}
G(x) \sim 1+ \frac{1}{\zeta(2)}(1-x) \ln (1-x).
\end{equation}
Using this we obtain
\begin{equation} \label{eq: N,tau,T asimp beta=1}
\begin{aligned}
& \nn \sim   \Gamma(1/2)\sqrt{2}\zeta(2) \frac{\sqrt{S}}{\ln S} ,
\\
& \tau \sim \frac{1}{\Gamma(1/2)\sqrt{2}\zeta(2)}\sqrt{S} \ln S ,
\\
& T \sim 2S.
\end{aligned}
\end{equation}

The conclusions are that in the asymptotic limit of large $S$ the behavior depends if the mean distance between food locations is finite or infinite. In the finite average case interestingly, while the food distribution does not affect the exponents of scaling relations, it does change the pre-factors, as follows, $\nn \sim fS^{1/2}$, $\tau \sim f^{-1} S^{1/2}$ and $T\sim S$ of Eq. \eqref{eq: N,tau,T asimp}, where $f=1/\av{l}$ is the food density. \\
However, food distribution with power-law tail, $P(l)\sim l^{-(1+\beta)}$, where $\beta<1$ (where $\av{l}$ diverges), yields exponents which depend on the distribution, i.e., $\nn \sim S^{\beta/2}$ and $\tau \sim S^{1-\beta/2}$ rather than $S^{1/2}$, while the scaling of $T\sim S^1$ is conserved. The pre-factor of $T$, however, depends on $\beta$, Eq. \eqref{eq: N,tau,T beta asimp}. Where $\beta=1$ a logarithmic correction appears, and $\nn \sim \sqrt{S}/\ln S$ and $\tau \sim \sqrt{S}\ln S$, Eq. \eqref{eq: N,tau,T asimp beta=1}.

\subsection{Examples of several distance distributions}

\begin{figure}[h]
	\centering
	\includegraphics[width=0.8\linewidth]{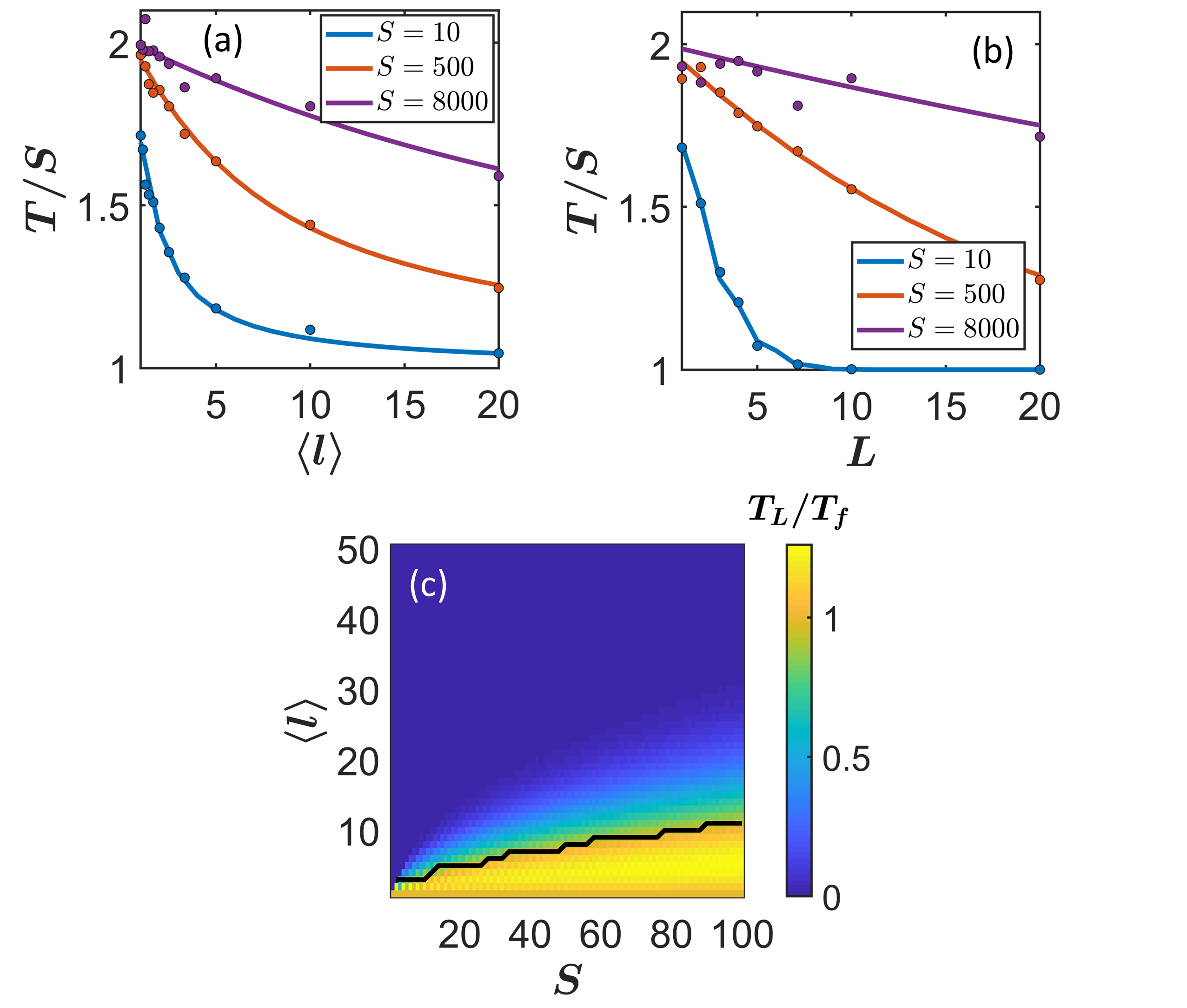}
	
	\caption{{\bf Random forager in one dimension. Comparison between constant distance between food and random food distributions.} (a) Results for random spread of food, from theory (lines), Eqs. (\ref{eq: F}), (\ref{eq: T}) and (\ref{eq: G random}), and simulations (symbols) averaged over $10^3$ realizations, show good agreement. (b) Results for a constant distance between food units. Lines represent the theory, Eqs. (\ref{eq: F}), (\ref{eq: T}) and (\ref{eq: G const}),  and symbols are 
	simulation results. Good agreement can be observed. (c) Phase diagram that compares which food distribution for the same $\av{l}$ (same amount of food) is better and yields longer lifetime of the forager. Above the black line the random spread gives longer life ($T_f$) while below the line, where $S$ is significantly larger than $\av{l}$, the constant distance enables longer lifetime ($T_L$). In (c) the life times are calculated excluding the last walk which is always $S$ steps for any food distribution.}
\label{fig: random const comparison}
\end{figure}

\noindent
\textbf{I. Uniform distance between food locations} \\
Here we consider the case where $l$, the distance between food locations, is uniform, $l=L$. Namely, 
\begin{equation}
P(l)=\delta_{l,L}.
\end{equation}
In this case
\begin{equation} \label{eq: G const}
	G(x)=\sum_{l=1}^{\infty} \delta_{l,L}x^l=x^L.
\end{equation}
Note that when $L=1$, then $G(x)=x$, and we recover the case of food is everywhere \cite{bhat2017starvation}.\\
Substituting Eq. \eqref{eq: G const} in Eq. \eqref{eq: F}, we have the theory for the constant distance between food, shown in Fig. \ref{fig: random const comparison}. \\
The scaling for large $S$ is, according to Eq. \eqref{eq: N,tau,T asimp}, $T\sim S, \ \nn \sim L^{-1}S^{1/2}, \ \tau \sim L S^{1/2}$.\\

\begin{figure}[h]
	\centering	
	\includegraphics[width=1\linewidth]{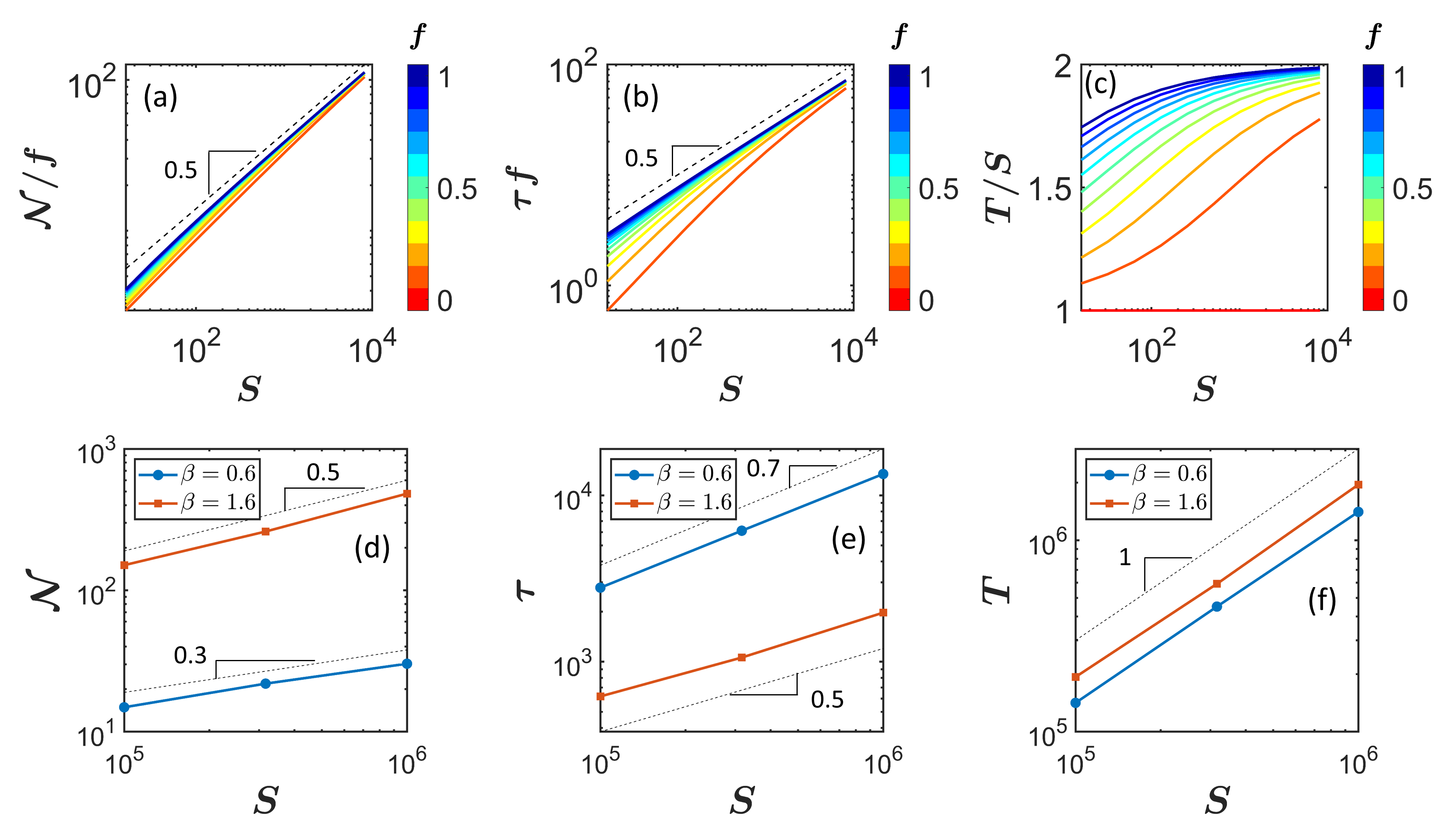}
	\caption{{\bf Random forager in one dimension with random (uniform) and power-law food distributions.} In the upper panels we show results of theory for random uniform spread of food, Eqs. \eqref{eq: F},\eqref{eq: N},\eqref{eq: tau},\eqref{eq: T} and \eqref{eq: G random}. One can see that the scaling relations for large $S$ in Eq. \eqref{eq: N,tau,T asimp} are valid, (a) $\nn\sim fS^{1/2}$ , (b) $\tau \sim f^{-1}S^{1/2}$ , and (c) $T\sim S $. Analysis of the behavior for small values of $f$ is presented in the Appendix in Fig. \ref{fig: random dist app}. In the lower panels, (d,e,f), we show results of simulations for power-law distribution of distance between food, $P(l)=Al^{-(1+\beta)}$. The results agree well with the theoretical scaling (dashed lines) found in Eq. \eqref{eq: N,tau,T beta asimp}, $\nn \sim S^{\beta/2}$, $\tau \sim S^{1-\beta/2}$, $T \sim S$, for $\beta<1$, and with the scaling in Eq. \eqref{eq: N,tau,T asimp}, $\nn\sim S^{1/2}, \ \tau\sim S^{1/2}, \ T \sim S$, for $\beta>1$. The results of simulations have been averaged over $10^3$ realizations. Note that in (f) for $\beta=0.6$ the factor $T/S$ is close to 1.5 instead of 1.99 as found in Eq. \eqref{eq: N,tau,T beta asimp}. This is since the analytical approximation in Eq. \eqref{eq: N,tau,T beta asimp} should be valid only for much larger S.
	}
	\label{fig: random dist}
\end{figure}

\noindent
\textbf{II. Random spread of food - likelihood $f$ of having food in each site} \\
Here we assume that randomly each site has food with probability $f$. Hence, the chance that an arbitrary food unit has, at a certain direction, the closest food at distance $l$, is  
\begin{equation} \label{eq: p(l) random}
P(l) = f(1-f)^{l-1}.
\end{equation}
Thus, Eq. \eqref{eq: p(l) random} is the distance distribution between food.
The average distance is related to the density by $\av{l}=1/f$.
Thus,
\begin{equation} \label{eq: G random}
G(x) = \sum_{l=1}^{\infty} f(1-f)^{l-1}  x^l = \frac{fx}{1-(1-f)x}.
\end{equation}
Note that when $f=1$, then $G(x)=x$, and we recover the case of food is everywhere \cite{bhat2017starvation}.\\
Substituting Eq. \eqref{eq: G random} in Eq. \eqref{eq: F} provides the theory for random spread of food, shown in Figs. \ref{fig: random const comparison} and \ref{fig: random dist}. \\
The scaling for large $S$ is, according to Eq. \eqref{eq: N,tau,T asimp}, $T\sim S, \ \nn \sim fS^{1/2}, \ \tau \sim f^{-1}S^{1/2}$.\\

For this food distribution, Eq. \eqref{eq: p(l) random}, we analyze in Appendix \ref{sec: app small f} also the behavior of $T,\tau,N$ in the limit of small $f$ for given $S$, and we get $T \sim f$, $\nn \sim f $ and $\tau \sim Constant$, see Fig. \ref{fig: random dist app}. \\

In Fig. \ref{fig: random const comparison} we also study which way to spread the food in space is better for the forager to live longer. Given the same amount of food we compare the results of life time between random spread of food and a constant distance between food. The black line in the phase diagram, Fig. \ref{fig: random const comparison}c, distinguishes between the two cases. For parameters below this line it is better to have a constant distance while above this line random distribution of food increases the life time of the forager. One can see that if $S \gg \av{l}$ then constant distance between food leads to a longer life time because random distribution will create at some place a long gap which causes starving. On the other hand, if $\av{l} \approx S$ then random spread is better, because $l$ might be many times lower than $\av{l}\approx S$, and thus the forager will probably get to cross this desert, unlike in constant distance $L \approx S$ where the forager will starve very fast. \\


\noindent
\textbf{III. Power law distribution of distances between food units} \\
If the spread of food is uniformly random, the distances between food are distributed exponentially as shown above, Eq. \eqref{eq: p(l) random}. In reality in many cases food is clustered such that most distances are short but few are long, what can be described by a power law distribution of distances between food locations. Therefore, we assume now that $P(l)$ fulfills
\begin{equation}
P(l) = A l^{-(1+\beta)},
\end{equation}
where $A=1/\zeta(1+\beta)$, and $\zeta$ is Riemann zeta function.\\
In this case the generating function is,
\begin{equation} \label{eq: G power law}
G(x) = A \sum_{l=1}^{\infty} l^{-(1+\beta)} x^l = \frac{{\rm Li}_{1+\beta}(x)}{\zeta(1+\beta)},
\end{equation}
where ${\rm Li}_{1+\beta}(x)$ is the polylogarithm of order $1+\beta$.
Substituting Eq. \eqref{eq: G power law} into Eq. \eqref{eq: F} provides the theory for a power law distribution of distances between food units.

For $\beta>1$ follows $\av{l}<\infty$, and for $\beta\leq1$ follows $\av{l}=\infty$. We analyzed above both cases for the asymptotic behavior for large $S$ yielding Eqs. \eqref{eq: N,tau,T asimp} and \eqref{eq: N,tau,T beta asimp}. For $\beta>1$, the average is finite, and the scaling is $\nn\sim S^{1/2}$,$\tau \sim S^{1/2}$ and $ T \sim S$, while for $\beta<1$, $\nn \sim S^{\beta/2}, \ \tau \sim S^{1-\beta/2}, \ T \sim S$. For $\beta=1$, we obtain logarithmic corrections to the scaling, $\nn \sim \sqrt{S}/\ln S$, $\tau \sim \sqrt{S}\ln S$ and $T\sim S$, Eq. \eqref{eq: N,tau,T asimp beta=1}. Results for this power law distribution and the scaling relations are presented in Fig. \ref{fig: random dist}.


\section{One dimension - Smelling forager} \label{sec: smell 1d}

In this chapter we study the case where each unit of food generates a smell felt by the forager and direct him towards the food. We assume that the smell decays with the distance from its source. All smell to the forager's right is summed up to $F_R$, and all smell to the left, to $F_L$. Then, the probability to go right, $p_R$, or left, $p_L$, is determined according to $F_R$ and $F_L$ simply by
\begin{equation}
	p_{R,L} = \frac{F_{R,L}}{F_R+F_L}.
\end{equation}
Food is distributed all over a one dimensional lattice, with some distance distribution $P(l)$ between food locations.
Here, given $P(l)$, we focus on the question whether the forager has a non-zero probability to live forever, $p_{\infty}$, or it is certainly mortal. To study this question we analyze two decay functions of smell, power law decay and exponential decay.

\subsection{Power Law decay of smell} \label{sec: power law smell}

Here we assume the decay of smell with distance is according to $d^{-\alpha}$, where $d$ is the distance between the locations of the forager and the food units which are the sources of smell. Note that if $\alpha\leq 1$ the total smell to each side diverges, and thus the forager walks completely randomly, a case that has been discussed above. Therefore we consider here only the case $\alpha>1$ and investigate the impact of smell.

We want to explore whether immortality exists. The reason that immortality might be possible is that as long as the forager propagates in one direction, its bias to this direction gets stronger because of the effect of smell. The question is if and in which conditions, this intensification is significant enough, and forager would live forever. \\
We define $P(l)$ to be the distribution of distances between food units locations.
In order to explore immortality, we treat separately two cases: (i) the original distance between food cannot be larger than $S$ according to $P(l)$, and (ii) the distance between food can be longer than $S$ according to $P(l)$. 
\\


\subsubsection*{(i) The case where the distance between food cannot be larger than $S$}

For this case, we find that there is an immortality phase which is dependent on the value of $\alpha$. There is a critical value $\alpha_c$ below which the forager will die at finite time with probability 1, and above which there is a nonzero chance to live forever. This $\alpha_c$, as we will show, depends on the distribution of food. In order to find $\alpha_c$ , we follow the steps in \cite{sanhedrai2019epl} and adjust them to our model as follows. \\
First, we define some useful quantities. $p_{\infty}$ is the probability to live forever. $\phi(D)$ is the chance to get the next meal, given the forager just ate and left behind a desert of size $D$ without food. $p_D$ is the probability to step towards the desert. We will focus on large $D$ because we are interested here in long time walks, which is needed for determining if the life time can be infinite. \\
Our goal is to determine if the probability to live forever, $p_{\infty}$, is zero. Since $p_{\infty}$ is the probability to always reach the next meal, hence
\begin{eqnarray} \label{eq: pinf}
p_{\infty} = \prod_{n=1}^{\infty} \phi(D_n),
\end{eqnarray}
where $n$ counts the meals, and $D_n$ is the size of the desert before the $n$th meal. Note that $D_{n+1} = D_{n} + l_{n} $ is satisfied where $l_{n}$ is distributed according to $P(l)$. \\
In order to find $\phi(D)$, we study first $p_D$.
After a long time of walking there is a large desert of size $D$ in one direction, thus the likelihood to step towards the desert is small and estimated \cite{sanhedrai2019epl} by
\begin{equation} \label{eq: pD}
p_D \sim D^{1-\alpha}.
\end{equation}
Next, we denote $\phi(D,l)$ as the likelihood to get a next meal given the next food is at distance $l$, and the desert on the other side is of size $D$. We consider long times for which $D$ is very large, hence $p_D$ is small, and thus the chance to starve $1-\phi(D,l)$ is small. Its leading term comes from the possibility with minimum number of steps, $k$, towards the desert among $S$ steps, such that the forager does not get the next food. This $k$, in our model with food distribution, depends on $l$, thus we denote it by $k_l$. 
It was shown in \cite{sanhedrai2019epl} that the chance not to escape a desert is
\begin{equation} \label{eq: phi(D,l)}
1-\phi(D,l) \sim  p_D^{k_l}.
\end{equation}
Here, $k_l$ satisfies
\begin{equation}
(S-k_l) -k_l \leq l-1,
\end{equation}
or
\begin{equation}
k_l \geq \frac{S-l+1}{2}.
\end{equation}
Because $k_l$ is minimal,
\begin{equation} \label{eq: kl}
k_l = \left \lceil \frac{S-l+1}{2} \right \rceil .
\end{equation}
The next step is to find $\phi(D)$, the desert escape probability without knowing the distance $l$ from the next food. 
We denote $l^*$ as the maximal possible distance between food according to $P(l)$. In this section, $l^* \leq S$ because $\Pr(l>S)=0$. \\
Then, the likelihood to get a next meal, $\phi(D)$, where $l$ is not given, using Eq. \eqref{eq: phi(D,l)}, is
\begin{equation}
1-\phi(D) = \sum_{l=1}^{\infty} [1-\phi(D,l)] P(l) = 
\sum_{l=1}^{l^*} [1-\phi(D,l)] P(l) 
\sim  \sum_{l=1}^{l^*} p_{D}^{k_l} P(l).
\end{equation}
Because $p_D$ is very small, the dominant term in the sum is the one with the minimal exponent $k_l$, which is for the largest $l$, i.e., $l^*$. Thus, recalling Eq. \eqref{eq: pD}, we get
\begin{equation} \label{eq: phi(D)}
1-\phi(D) \sim p_{D}^{k_{l^*}} \sim D^{(1-\alpha)k_{l^*}} .
\end{equation}
Now we can evaluate $p_{\infty}$ using Eqs. (\ref{eq: pinf}) and (\ref{eq: phi(D)}),
\begin{eqnarray} \label{eq: pinf convergence}
p_{\infty} = \prod_{n=1}^{\infty} \phi(D_n) = \exp \left( \sum_{n=1}^{\infty} \ln ( \phi(D_n)) \right) \sim  \exp \left( -\sum_{n=1}^{\infty} D_n^{(1-\alpha)k_{l^*}} \right).
\end{eqnarray}
Thus, $p_{\infty}=0$ if and only if the sum in the exponent diverges. Because the differences between $D_n$ are bounded by $S$, the sum diverges simply when,
\begin{eqnarray}
(1-\alpha)k_{l^*} \geq -1
\end{eqnarray}
Thus,
\begin{equation} \label{eq: ac}
\alpha_c = 1+ \frac{1}{k_{l^*}}  = 1 + \frac{1}{\left \lceil \frac{S-l^*+1}{2} \right \rceil} ,
\end{equation}
and $p_{\infty}=0$ if $\alpha\leq \alpha_c$, i.e., the forager will definitely die  at finite time, while for $\alpha > \alpha_c$ there is a non-zero chance to survive forever. \\
This result is not trivial because the naive guess might be that $\alpha_c$ should be determined by the average distance, $\av{l}$, however we find that the maximal distance, $l^*$, is the quantity determining $\alpha_c$.\\
For the simple case where the distance between food is constant, $l=L \leq S$, the result is simply
\begin{equation} \label{eq: ac const}
\alpha_c = 1 + \frac{1}{\left \lceil \frac{S-L+1}{2} \right \rceil} .
\end{equation}
Note that while in general critical exponents are not sensitive to microscopic characteristics and change only with dimension or symmetry changes \cite{stanley1971phase,bunde1991fractals}, here the critical exponent is governed both by a quantitative feature of the forager ($S$) and by a quantitative feature of the spread of food in space ($L$).  \\ 

\begin{figure}[h]
	\centering
	\includegraphics[width=0.99\linewidth]{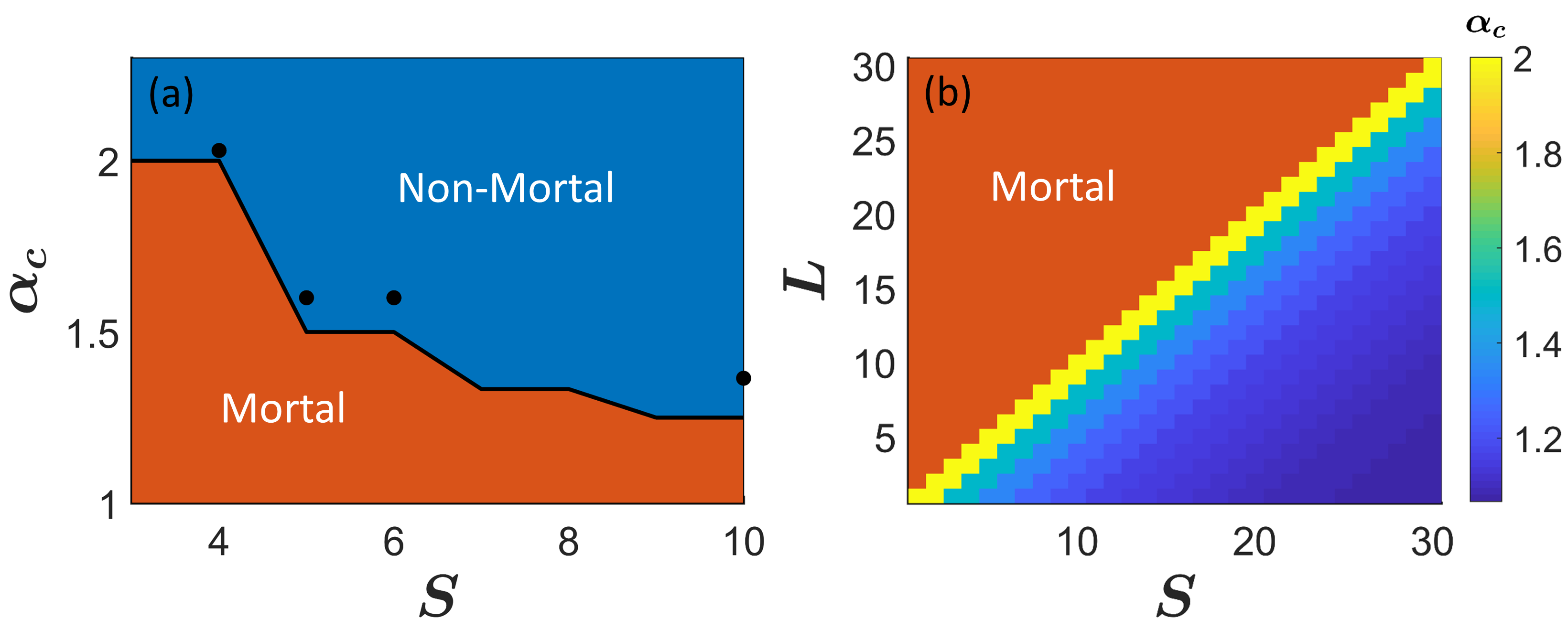}
	\caption{{\bf Smelling forager with a constant distance between food and power law decay of smell.} (a) Results of theory (line between two phases) and simulations (symbols) for forager in one dimension with power law decay of smell, and food is distributed with a constant distance, $L=3$, between food units. The theory is taken from Eq. \eqref{eq: ac const}, and the simulations have been performed over $10^8$ realizations, where $\alpha_c$ is determined by the maximal value of $\alpha$ for which there was no forager which lived forever in any realization, i.e., in all $10^8$ realizations the forager died. The deviation between theory and simulations is reasonable because the theory finds when the probability to live forever is completely zero, whereas the simulations find when the probability is small enough such that it does not appear in the finite number of realizations, and thus it happens for a slightly larger value of $\alpha$. (b) Shows the dependence of $\alpha_c$ on $L$ and $S$ according to Eq. \eqref{eq: ac const}. Where $L>S$ of course the forager dies after one walk and it is mortal for any value of $\alpha$. The color represents the value of $\alpha_c$ required for immortality which changes with $L$ and $S$.}
	\label{fig: ac smell constant}
\end{figure}

In Fig. \ref{fig: ac smell constant}a we show the results for $\alpha_c$ of theory and simulations for a constant distance between food, $L=3$. Fig. \ref{fig: ac smell constant}b shows the result of Eq. \eqref{eq: ac const}. Of course where $L>S$ the forager is mortal, however, for $L\leq S$ each point has a critical value of the exponent $\alpha$ above which the forager is immortal.\\


\subsubsection*{(ii) The case where the distance between food units can be larger than $S$}

At this scenario, we show that there is no chance to live forever because after each eating there is a nonzero probability that $l>S$, and when this happens the forager will certainly die. Hence there is no immortality phase, and the lifetime $T$ is finite for any $\alpha$. Where $\alpha$ is large such that the forager walks almost certainly towards the closest food, we can find the lifetime $T$.\\

First, we prove that the forager is mortal in this case.
Let us observe the forager after creating a desert larger than $S$. After each meal the likelihood to eat again is $\phi(D)$. The distance to next food, $l$, is random and sampled from $P(l)$. If $l>S$ it will not eat again for sure. It is easy to see that $\phi(D) \leq 1-\Pr(l>S)$, and that $\Pr(l>S)$ is nonzero, and independent on $D$ or on time. \\ 
Then, we approach to find $p_{\infty}$, the chance to live forever, according to Eq. \eqref{eq: pinf},
\begin{eqnarray}
p_{\infty} = \prod_{n=1}^{\infty} \phi(D_n) \leq \prod_{n=1}^{\infty} \left[1-\Pr(l>S)\right]=0 .
\end{eqnarray}
Namely, the forager will certainly die in a finite time for any value of $S$ and $\alpha$.\\

Since the lifetime is finite, we wish to calculate the average number of meals, $\nn$, for large $\alpha$. We assume that $\alpha$ is large such that the forager steps always towards closest food. Hence, the only chance to die is if $l>S$. Therefore, the chance to reach the next meal is $\phi = 1-\Pr(l>S)$, and from the average of geometric distribution follows,
\begin{eqnarray} \label{eq: N large alpha}
\mathcal{N} = \frac{1}{\Pr(l>S)}. 
\end{eqnarray}
Next, we study the scaling derived from Eq. \eqref{eq: N large alpha} for two distance distributions discussed above, random and power law. \\

\noindent
{\bf I. Random spread of food in space}\\
In this case we assume there is a likelihood $f$ of having food in each site. The result is that the distribution of distance between food is geometrical,
\begin{equation}
	P(l)=f(1-f)^{l-1}.
\end{equation}
It is clear that $\Pr(l>S)>0$.
Thus, there is no immortality regime. Let us find $\Pr(l>S)$,
\begin{equation}
\Pr  (l>S)  = \sum_{l=S+1}^{\infty}P(l) 
= f\sum_{l=S+1}^{\infty}(1-f)^{l-1} = (1-f)^{S}.
\end{equation}
Therefore, based on Eq. \eqref{eq: N large alpha}, $\nn$ for large $\alpha$ is,
\begin{eqnarray}
	\nn = \left[\frac{1}{(1-f)}\right]^{S}.
\end{eqnarray}
The average time between meals is smaller than $\av{l}$ because it is an average given $l\leq S$. However it is in the order of magnitude of $\av{l}$. Therefore, $T=\nn \tau+S$ obeys the same scaling as $\nn$. \\
Thus, for large $\alpha$, 
\begin{equation}\label{eq: T exp large alpha}
T \sim [1/(1-f)]^S,
\end{equation}
namely the mean lifetime increases exponentially with $S$.
\\

\noindent
{\bf II. Power law distribution of distances between food units}\\
Here, we assume that $P(l)$ satisfies
\begin{equation}
P(l) = A l^{-(1+\beta)}.
\end{equation}
In this case,
\begin{eqnarray}
\Pr(l>S) = 
\sum_{l=S+1}^{\infty} Al^{-(1+\beta)} \approx A \int_{S+1}^{\infty} l^{-(1+\beta)} \dd l =\frac{A}{\beta} (S+1)^{-\beta}.
\end{eqnarray}
Therefore, plugging this in Eq. \eqref{eq: N large alpha},
\begin{equation}
	\nn \sim S^{\beta}.
\end{equation}
Here $\tau$ might be dependent strongly on $S$ because the tail is not neglected, so it matters where it is cut, 
\begin{gather}
\tau = \sum_{l=1}^{S}l A l^{-(1+\beta)} \approx A \int_{1}^{S} l^{-\beta} \dd l = \frac{A}{1-\beta} \left( S^{1-\beta}-1 \right)
\sim
\begin{cases}
S^{1-\beta}, & 0<\beta<1
\\
1, & \beta>1
\end{cases} .
\end{gather}
Therefore,
\begin{equation} \label{eq: T power law large alpha}
T = \nn \tau + S \sim 
\begin{cases}
S, & 0<\beta<1
\\
S^{\beta}, & \beta>1
\end{cases} .
\end{equation}

\subsubsection*{Summary of all cases}

\begin{figure}[h]
	\centering
	\includegraphics[width=1\textwidth]{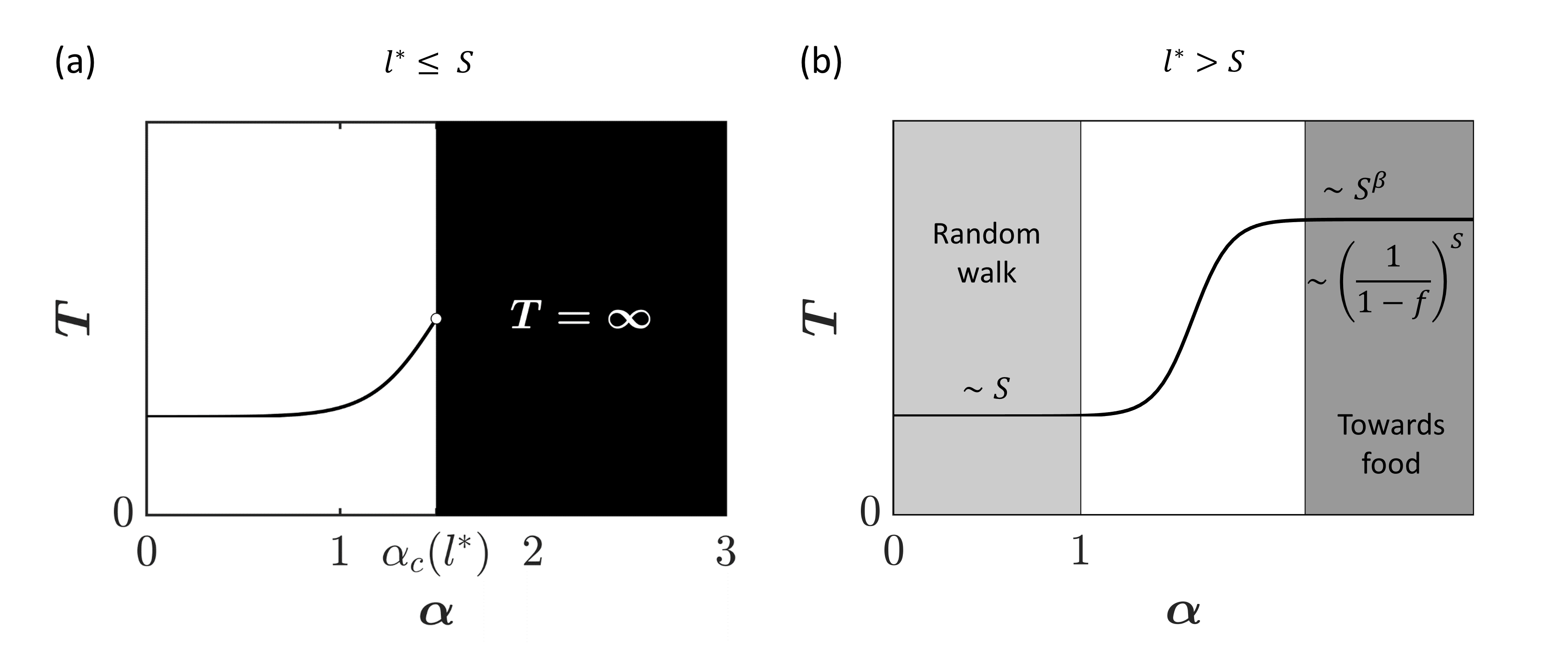}
	\caption{{ \bf Sketches of theory results for lifetime of smelling forager in one dimension} (a) If the maximal possible distance between food ($l^*$) is not greater than the starving time ($S$), then there is $\alpha_c$ above which the average lifetime is infinite. Its value depends on $l^*$ according to Eq. \eqref{eq: ac}. (b) If the distance between food can be larger than $S$, there is no immortality regime. However, the lifetime has a saturation for large $\alpha$, and in this region the scaling of $T(S)$ is found for power law distribution of distance between food, Eq. \eqref{eq: T power law large alpha}, and for random spread of food, Eq. \eqref{eq: T exp large alpha}. For $\alpha<1$ the total smell diverges, and hence the walk is random, thus the scaling is as in Eqs. \eqref{eq: N,tau,T asimp} and \eqref{eq: N,tau,T beta asimp}.
	}
	\label{fig: theory illustration}
\end{figure}

We denote $l^*$ as the maximal $l$ with non-zero probability. When $l^*\leq S$, then there is $\alpha_c$ above which $p_{\infty}>0$, therefore $T=\infty$, and $\alpha_c$ depends on the distance distribution as $\alpha_c=1+1/\lceil (S-l^*+1)/2 \rceil$. Therefore, $T(\alpha)$ is an increasing function that diverges at $\alpha_c$ as illustrated in Fig. \ref{fig: theory illustration}a. \\
When $l^*>S$, then $T<\infty$ for any $\alpha$.
Then we get that $T(\alpha)$ is an increasing function with $\alpha$, starting at the completely random case ($\alpha<1$) where the scaling is $T\sim S$ as in Eqs. (\ref{eq: N,tau,T asimp}) and (\ref{eq: N,tau,T beta asimp}) and approaching a saturation where $\alpha$ is large such that the forager always tends towards the closest food. Then the scaling is a power law or exponential as in Eqs. (\ref{eq: T exp large alpha}) and (\ref{eq: T power law large alpha}). See Fig. \ref{fig: theory illustration}b.

\subsection{Exponential decay of smell}

Here we assume the decay of smell with distance is according to $\exp (-\lambda d)$. The results for this case can be studied using the same formalism as in Sec. \ref{sec: power law smell} . \\
For food distribution that allows $l>S$, the forager is trivially mortal. The analysis of Eqs. (\ref{eq: T exp large alpha}) and (\ref{eq: T power law large alpha}) is valid for large $\lambda$ the same as for large $\alpha$. \\
For food distribution where all $l<S$, similar steps as in power law decay can be performed and obtain Eq. \eqref{eq: pinf convergence}. Then for exponential decay of smell, the sum in the exponent is exponential, therefore it converges for any $\lambda>0$. Thus, in contrast to Eq. \eqref{eq: ac} where we get critical $\alpha_c$, in the case of exponential decay of smell there is no critical $\lambda$, and $p_{\infty}>0$ for any $\lambda>0$, and the mortal regime vanishes.

\section{Forager in two dimensions}
\FloatBarrier
In this chapter we analyze a forager walking in a two dimensional lattice. We consider several types of walk and compare between them, random walk, short range smell (the forager detects only sites in distance one), long range smell, and complete bias towards smell. We also consider several distributions of food in space, food is everywhere, food is located in constant distances, and random spread of food with density $f$.

\subsection{Space is full of food}

\begin{figure}
	\includegraphics[width=0.95\textwidth]{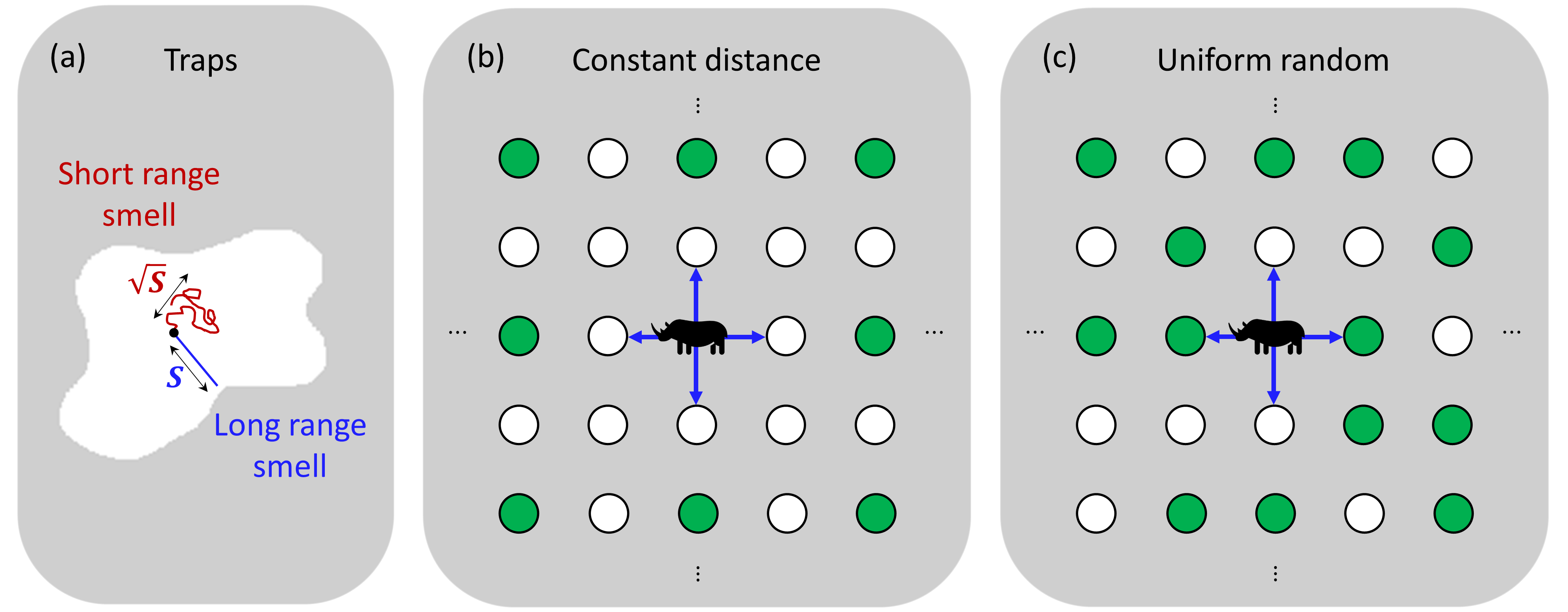}
	\caption{{\bf Illustrations of forager in two dimensions.} (a) The forager dies when it creates a loop and goes inside. The loop should be large enough such that after it eats most of the food inside and  find itself at the middle of desert, it does not succeed escaping the desert in $S$ steps. At this situation, long range smelling forager and one with short range are very different but can be mapped. While the long range smelling forager goes directly towards the closest food (blue straight line), the short range one walks randomly (red random path). Typically random walker reaches distance of $\sim \sqrt{S}$ in $S$ steps, while the direct walk reaches a distance $S$. Hence the short range smelling forager should have starving time of $S^2$ to die at the same trap as the long range smeller forager with starving time of $S$. Therefore, we obtain Eq. \eqref{eq: short-long relation}. (b) Illustration of a constant distance between food in 2D Lattice where $L=2$. The filled circles represent food, while the white empty ones represent empty sites. Theory for this case is given in  Eqs. \eqref{eq: full spread relation} and \eqref{eq: T spread 2d}. Simulation results of this case are shown in Fig. \ref{fig: 2D food density}. (c) Illustration of uniform random spread of food in two dimensions with density $f\approx 1/2$. Simulation results for this scenario are presented in Fig. \ref{fig: 2D food density}.
	}
	\label{fig: 2d illustration}
\end{figure}

We define a forager with short range smell as one that if there is food in a site next to it, it steps towards food with probability 1. However the forager does not consider food that are at distances more then 1. Such a forager has been investigated in \cite{bhat2017does, bhat2017starvation},  and it was shown that it dies because of traps it creates to itself, i.e., when the forager closes a loop, it might go inside at the next step, and then eat all food inside, until it finds itself at the middle of a desert without food which it created. Then, since there is no close food it walks randomly. If the loop of the trap is large enough the forager might starve before it reaches the edge of its self made desert, see Fig. \ref{fig: 2d illustration}a. \\
In contrast, a smelling forager senses also far food. Let us consider a forager that steps with probability 1 to the direction of the closest food. We call this forager \emph{perfect smelling forager}. This forager walks in 2D exactly the same as the short range smell forager we mentioned above, except that if it finds itself in a middle of a desert it does not walk randomly but walks certainly towards the closest food, see Fig. \ref{fig: 2d illustration}a. \\
We next consider the relation between these two cases, short range smelling and perfect smelling. We argue, using a rigorous mapping, that perfect smelling with starving time $S$, is similar to a short range smelling forager with starving time $S^2$. The reason is that the short range smelling forager walks randomly inside the trap, and therefore reaches in $S$ steps a distance of order $\sqrt{S}$, while the perfect smelling forager moves in a straight line, thus reaches a distance $S$ in $S$ steps, see Fig. \ref{fig: 2d illustration}a. The conclusion is that if the function of lifetime of a short range smelling forager is known to be $T_{\rm short}(S)$, then for the lifetime of the perfect smelling forager,
\begin{equation} \label{eq: short-long relation}
	T_{\rm perfect} (S) \sim  T_{\rm short} (S^2).
\end{equation}

\begin{figure}[ht]
	\centering
	\includegraphics[width=0.95\textwidth]{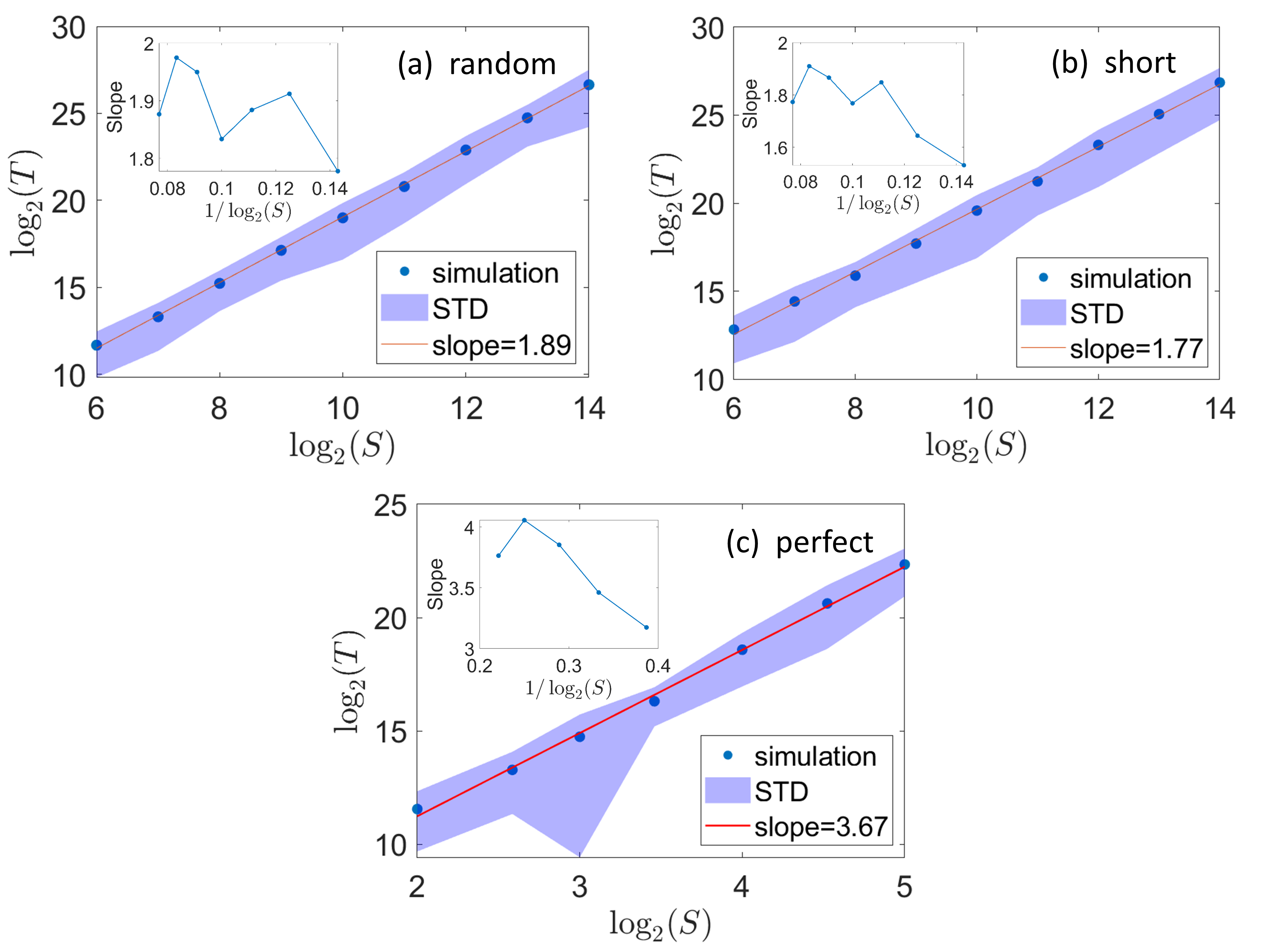}	
	\caption{{ \bf Lifetime in 2D where the space is full with food.} We show the scaling between lifetime, $T$, and starving time, $S$, at three cases: (a) random walk, (b) short range smell (steps towards food only if distance is one), and (c) perfect smelling forager (steps towards the closest food). It can be seen that for random forager and short range smelling forager the scaling is about $T \sim S^2$ when $S$ is getting large, while perfect smelling forager has scaling of about $T \sim S^4$. This confirms our theoretical argument that the transformation from short range smell to perfect smelling should be expressed by $S \to S^2$, see Fig. \ref{fig: 2d illustration}a and Eq. \eqref{eq: short-long relation}. The insets suggest that the exponents smaller than 2 and 4 are due to finite size systems, and when $S$ increases they reach the values 2 and 4 respectively.
	}
	\label{fig: 2D full}
\end{figure}

Computer simulations suggest, as presented in Fig. \ref{fig: 2D full}b, that for greedy forager with short range smell the mean life time scales similar to a random forager \cite{benichou2014depletion} approximately as
\begin{equation} \label{eq: T short}
T_{\rm short}\sim S^{2}.
\end{equation}
Thus, according to our prediction in Eq. \eqref{eq: short-long relation}, the mean lifetime of a forager with long range smell should scale approximately as 
\begin{equation} \label{eq: T long}
T_{\rm perfect} \sim S^4,
\end{equation}
and indeed this result is supported in Fig. \ref{fig: 2D full}c.\\
The meaning of Eqs. \eqref{eq: short-long relation},\eqref{eq: T short} and \eqref{eq: T long} is that the difference between short and long range of smell is dramatic, the exponent changes from 2 to 4 and the life time increases tremendously for perfect smelling forager. The result of Eq. \eqref{eq: T long} will serve us in the next chapter where we study forager with long range smell in two dimensions with food distribution in space.

\subsection{Space not full of food}

Here we consider a forager in 2D given some distribution of food in space. We focus on forager walking according to its sense of smell. We explore two distributions of food in space, constant distance $L$ between food, Fig. \ref{fig: 2d illustration}b, and random uniform spread of food with density $f$, Fig. \ref{fig: 2d illustration}c.

\subsubsection{Constant distance between food locations}

Let food located in 2D at points $(nL,mL)$ where $m$ and $n$ are all the integers, and $L$ is the distance between neighboring food units. The forager starts at point $(0,0)$ and its steps to right/left/up/down have size 1. We call this scenario a constant distance $L$ between food in 2D, see Fig. \ref{fig: 2d illustration}b.\\
Now, let us consider a smelling forager with a power law decay of smell, $d^{-\alpha}$, with large exponent $\alpha$, or exponential decay, $e^{-\lambda d}$, with large $\lambda$, namely the bias towards the food is very high and the forager walks almost always towards the closest food. We note that in this case, the walk is same as for constant distance $L=1$ (food is everywhere), except that each step now is replaced by $L$ straight steps. Therefore, 
\begin{equation} \label{eq: full spread relation}
	T_{\rm spread}(S,L)=L\cdot T_{\rm spread}(S/L,1)=LT_{\rm full}(S/L) ,
\end{equation}
where $T_{\rm full}$ is the life time of smelling forager in space full with food.
Thus, assuming a smelling forager in full space scales with $S$ as, 
\begin{equation}
T_{\rm full}(S) \sim S^{\gamma}, 
\end{equation}
then 
\begin{equation}
T_{\rm spread}(S,L) = L T_{\rm full} (S/L) \sim L (S/L)^{\gamma} = L^{1-\gamma}S^{\gamma},
\end{equation}
or in different shape, the scaling of $T$, $S$ and $L$, for perfect smelling forager in 2D with constant distance between food, is 
\begin{equation}
 \frac{T}{L}  \sim \left( \frac{S}{L} \right)^{\gamma} .
\end{equation}
This scaling is supported in Fig. \ref{fig: 2D food density}a where it can be seen that all points of different $S$ and $L$, where $S/L$ is large, lay on the same curve when plotting $T/L$ vs $S/L$, what validates the scaling we predicted theoretically in Eq. \eqref{eq: full spread relation}.
Using computer simulations we find that $T\sim S^{\gamma}$ for large $S$, and $\gamma \approx 4$. Hence, for a large ratio $S/L$, we expect
\begin{equation} \label{eq: T spread 2d}
	T_{\rm spread}(S,L) \sim L^{-3}S^4.
\end{equation}
In terms of density of food $f$, rather than the distance between food $L$, using the simple relation
\begin{equation*}
	f = 1/L^2,
\end{equation*}
we obtain the scaling
\begin{equation} \label{eq: T vs f 2d}
	T_{\rm spread}(S,f) \sim {f}^{\frac{\gamma-1}{2}}S^{\gamma} \approx {f}^{3/2}S^4.
\end{equation}
Note that this scaling is valid for large $S/L$, or, for large $S\sqrt{f}$, i.e., for $f \gg S^{-2}$. 


\subsubsection{Random uniform spread of food in space}

\begin{figure}[ht]
	\centering
	\includegraphics[width=0.95\textwidth]{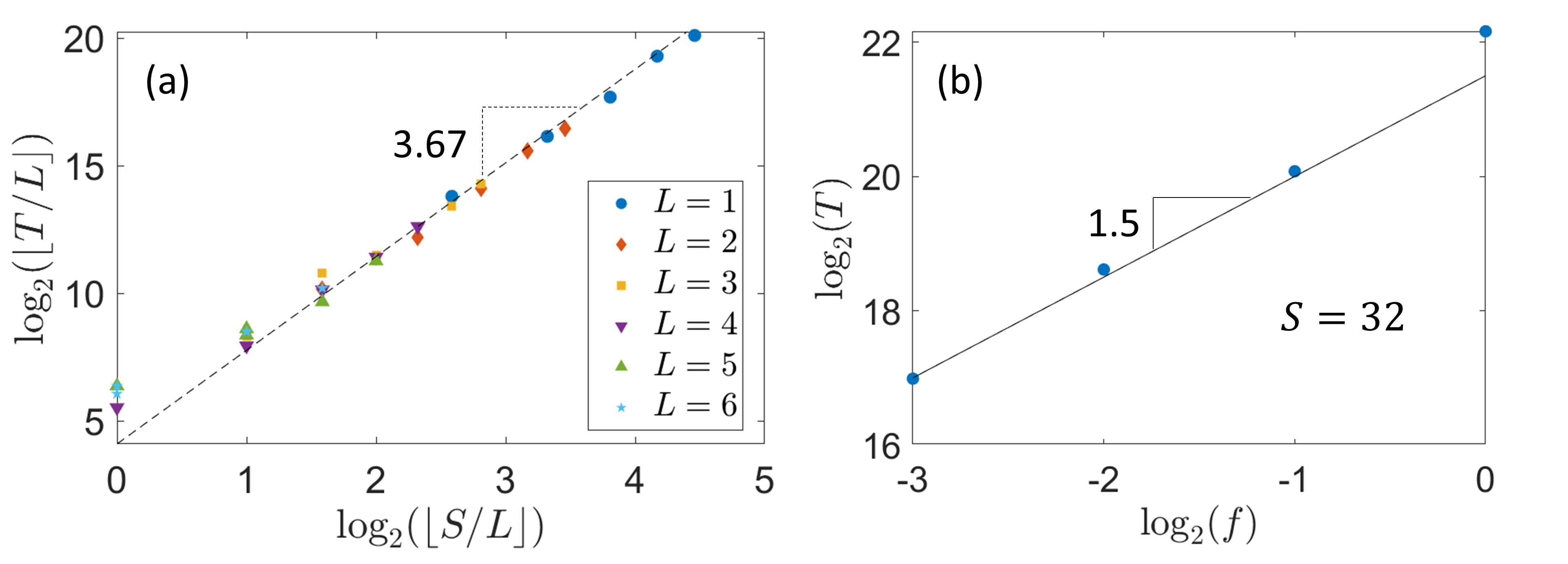}
	\caption{{ \bf Forager in two dimensions where space is not full with food.} (a) Here the distance between food units is constant, $L$, and the forager walks perfectly according to smell, i.e., towards the closest food. We show that the scaling with the distance $L$ is $T/L$ and $S/L$ when $S/L$ is large, hence all points lay approximately on one curve. This supports Eq. \eqref{eq: full spread relation} derived from theoretical considerations. This scaling allows us to obtain $T \sim L^{-3}$, see Eq. \eqref{eq: T spread 2d}. The dashed line represents the expected approximated slope according to the result of Fig. \ref{fig: 2D full}c. (b) Here food is distributed randomly with density $f$, see Fig \ref{fig: 2d illustration}c. The forager walks perfectly according to smell. We examine in this case the scaling of $T$ vs $f$. One can see that the approximated scaling in Eq. \eqref{eq: T vs f 2d}, $T\sim f^{3/2}$, found for constant distances, is approximately valid also for a uniform random  food distribution.  
	}
	\label{fig: 2D food density}
\end{figure}

Here we assume that at each site in two dimensional square lattice there is food with likelihood $f$. This probability $f$ is, therefore, the density of food, see the illustration in Fig. \ref{fig: 2d illustration}c. The forager walks according to a long range smell with high bias, such that it steps towards the closest food. In Fig. \ref{fig: 2D food density}b we show the results of the lifetime of such a forager for different values of density $f$. One can see that the approximated scaling we found for a constant distance between food in the previous section, Eq. \eqref{eq: T vs f 2d}, $T\sim f^{3/2}$, using the combination of simulations results and theoretical considerations, works well also for different cases of food distribution in space.

%
%

\section{Discussion}

We have studied a forager that walks in space where food is distributed in several fashions, a constant distance between food units, uniform random distribution and power law distribution of distances. We have considered foraging both in one and two dimensions. Moreover, we have treated a few types of forager's walk; random, according to short range smell and according to long range smell. We studied two cases of long-range smell. Smell decaying exponentially and as a power law. We found new scaling relations between forager's lifetime, number of meals, the starving time and the density of food in one and two dimensions. We also found how the immortality of a long range smelling forager in one dimension depends on the distribution of food in space.  

Further work could compare these results to experimental measurements, which also might lead to additional extensions to the model such as exploring cases incorporating the fact that food often appears in `patches' \cite{nevitt2008evidence}.  Likewise, multiple foragers living in the region could be considered with all of them depleting food sources \cite{martinez2013optimizing}. 

\FloatBarrier

\section{Acknowledgments}

We thank the Israel Science Foundation (Grant No. 189/19) and the joint China-Israel Science Foundation (Grant Bo. 3132/19), the BIU Center for Research in Applied Cryptography and Cyber Security, NSF-BSF Grant no. 2019740, and DTRA Grant no. HDTRA-1-19-1-0016 for financial support.

\bibliographystyle{unsrtnat}
\bibliography{Forager_refs}

\clearpage

\appendix
\setcounter{equation}{0}
\renewcommand{\theequation}{\thesection\arabic{equation}}

\section{Derivation of $\nn,\tau,T$ using $\mathcal{F}(z)$} \label{sec: app T,tau,N by F}

In this Appendix we summarize what relevant to us for derivation of $N$, $\tau$ and $T$ based on Ref. \cite{bhat2017starvation}.
After having the generating function $\mathcal{F}(z)$, we define ${E}(S)$ as the probability of a random forager to escape the desert, namely to get food before starving, given it starves after $S$ steps without food, and it just ate. One should note that
\begin{equation}
{E}(S) = \sum_{t=1}^{S} F(t),
\end{equation}
what implies regarding the generating functions
\begin{equation}
\mathcal{E}(z) = \frac{\mathcal{F}(z)}{1-z}.
\end{equation}
Next, we find the distribution of $N$, number of meals which is a geometric distribution,
\begin{equation}
p_{N } = (1-{E}) {E}^{N }.
\end{equation}
Hence for the average, $\nn  = \av{N}$,
\begin{equation}
\nn   = \frac{{E}}{1-{E}}
\end{equation}
Then, to evaluate $\tau$, we note that 
\begin{equation}
\tau = \frac{\sum_{t=1}^{S} t F(t)}{\sum_{t=1}^{S} F(t)} =  \frac{\sum_{t=1}^{S} t F(t)}{{E}(S)} \equiv \frac{\pi(S)}{{E}(S)}.
\end{equation}
Therefore, we approach to find $\pi(S)$ via its generating function, which obeys
\begin{align}
\Pi(z) = \frac{z  \mathcal{F}'(z)}{1-z}.
\end{align}
Finally, for the lifetime of the forager we obtain
\begin{gather}
T(S) = \tau \nn   + S = \frac{\pi(S)}{E(S)} \frac{E(S)}{1-E(S)} +S  = \frac{\pi(S)}{1-E(S)} + S .
\end{gather}
To summarize, given the distribution of food in space $P(l)$, we find $G(x)$, what provides $\mathcal{F}(z)$. Using the last one, we find $\nn, \tau, T$.

\section{Asymptotic behavior where S is large in one dimension} \label{sec: app large S}
We analyze for general distance distribution between food $P(l)$ the asymptotic behavior of $T, \nn$ and $\tau$ for large $S$. For this goal we observe the limit $z\to1$ and use the Tauberian theorems \cite{feller-vol-2}.  $\alpha(z)$ fulfills $\alpha(1)=1$. Close to 1 $\alpha(z) \sim 1- \sqrt{2}\sqrt{1-z}$. In cases for which the mean distance, $\av{l}$, is finite, $G'(1)=\av{l}$. In addition, $G(1)=1$. Therefore, for $x\to1$
\begin{gather}
G(x) \sim G(1)+G'(1)(x-1) = 1 - \av{l}(1-x) .
\end{gather}
Hence, for $z\to1$
\begin{align}
\mathcal{F}(z) = G(\alpha(z)) \sim 1-\av{l}(1-\alpha(z))
\sim 1-\av{l}\sqrt{2}\sqrt{1-z} .
\end{align}
Hence,
\begin{gather}
\mathcal{E}(z) = \frac{\mathcal{F}(z)}{1-z} \sim  \frac{1}{1-z} -\frac{\sqrt{2}\av{l}}{\sqrt{1-z}} .
\end{gather}
Thus,
\begin{equation}
E(S) \sim  1- \frac{\sqrt{2}\av{l}}{\Gamma(1/2)} \frac{1}{\sqrt{S}} = 1 - \frac{\sqrt{2}\av{l}}{\sqrt{\pi}} \frac{1}{\sqrt{S}}.
\end{equation}
Hence,
\begin{equation}
\nn   = \frac{E}{1-E} \sim   \sqrt{\frac{\pi}{2}} \frac{1}{\av{l}} \sqrt{S} .
\end{equation}
In addition,
\begin{align}
\Pi(z) = \frac{z\mathcal{F}'(z)}{1-z} \sim \frac{\sqrt{2}\av{l}}{2(1-z)^{3/2}}.
\end{align}
Therefore,
\begin{align}
\pi(S) \sim \frac{\sqrt{2}\av{l}}{2\Gamma(3/2)}\sqrt{S} = \frac{\sqrt{2}\av{l}}{\sqrt{\pi}}\sqrt{S}.
\end{align}
And thus,
\begin{equation}
\tau = \frac{\pi}{E} \sim \frac{\sqrt{2}\av{l}}{\sqrt{\pi}}\sqrt{S}.
\end{equation}
Then,
\begin{align}
T(S) =\frac{ \pi(S)}{1-E(S)}+S\sim \frac{(\sqrt{2}\av{l}/\sqrt{\pi})\sqrt{S}}{(\sqrt{2}\av{l}/\sqrt{\pi})/\sqrt{S}} +S = 2S.
\end{align}
\\

\noindent
{\bf Power-law distance distribution}\\
Here we consider power law distance distribution $P(l) = A l^{-(1+\beta)}$ between food units, where $A=1/\zeta(1+\beta)$, and $\zeta$ is Riemann zeta function. The generating function of this distribution is,
\begin{equation}
G(x) = A \sum_{l=1}^{\infty} l^{-(1+\beta)} x^l = \frac{{\rm Li}_{1+\beta}(x)}{\zeta(1+\beta)},
\end{equation}
where ${\rm Li}_{1+\beta}(x)$ is the polylogarithm of order $1+\beta$.
Here $\av{l}$ is not finite in all cases, and one should separate the treatment into two cases. 
The expansion of polylogarithm around 1 is
\begin{equation}
{\rm Li}_{1+\beta}(x) \sim
\begin{cases}
\zeta(1+\beta) - \zeta(\beta)(1-x) , & \beta>1
\\
\zeta(1+\beta) + \Gamma(-\beta)(1-x)^{\beta} , & 0<\beta<1
\\
\zeta(2) + (1-x) \ln (1-x) , & \beta=1
\end{cases}.
\end{equation}
For $\beta>1$, the mean distance is finite, and we already obtained the asymptotic behavior in this case above.\\
For $\beta<1$ we analyze the asymptotic behavior in large $S$. To this end, we expand the relevant functions in the limit $z\to1$.
At $z\to 1$, $\alpha(z) \sim 1 - \sqrt{2}\sqrt{1-z}$, and therefore
\begin{gather}
\mathcal{F}(z) = G(\alpha(z)) =\frac{ {\rm Li}_{1+\beta}(\alpha(z))}{\zeta(1+\beta)} \sim 
1 + \frac{\Gamma(-\beta)}{\zeta(1+\beta) } (2(1-z))^{\beta/2} \sim 
1- C (1-z)^{\beta/2}.
\end{gather}
Next, for the generating function of $E(S)$,
\begin{gather}
\mathcal{E}(z) = \frac{\mathcal{F}(z)}{1-z}\sim  \frac{1}{1-z} - C (1-z)^{\beta/2-1} .
\end{gather}
Thus,
\begin{equation}
E(S) \sim  1- \frac{C}{\Gamma(\beta/2)} S^{-\beta/2} .
\end{equation}
Hence,
\begin{equation}
\nn   = \frac{E}{1-E} \sim   \frac{\Gamma(\beta/2)}{C} S^{\beta/2} .
\end{equation}
In addition, for the generating function of $\pi(S)$
\begin{align}
\Pi(z) = \frac{z\mathcal{F}'(z)}{1-z} \sim \frac{C\beta}{2(1-z)^{2-\beta/2}}.
\end{align}
Therefore,
\begin{align}
\pi(S) \sim \frac{C\beta}{2\Gamma(2-\beta/2)}S^{1-\beta/2} .
\end{align}
Then
\begin{align}
T(S) =\frac{ \pi(S)}{1-E(S)}+S\sim \frac{C\beta}{2\Gamma(2-\beta/2)}S^{1-\beta/2} \frac{\Gamma(\beta/2)}{C} S^{\beta/2}   +S = \left(\frac{\Gamma(1+\beta/2)}{\Gamma(2-\beta/2)}+1\right)S.
\end{align}
\\
In the edge case  $\beta=1$, we get at $z\to1$
\begin{gather}
\mathcal{F}(z) = G(\alpha(z)) =\frac{ {\rm Li}_{2}(\alpha(z))}{\zeta(2)} \sim 
1 + \frac{1}{\zeta(2) } \sqrt{2(1-z)}\ln\sqrt{2(1-z)}  \sim 
1 + C \sqrt{1-z}\ln(1-z).
\end{gather}
Next, for the generating function of $E(S)$,
\begin{gather}
\mathcal{E}(z) = \frac{\mathcal{F}(z)}{1-z}\sim  \frac{1}{1-z} + C \frac{\ln(1-z) }{\sqrt{1-z}}.
\end{gather}
Thus,
\begin{equation}
E(S) \sim  1 - \frac{C}{\Gamma(1/2)} \frac{\ln S}{\sqrt{S}} .
\end{equation}
Hence,
\begin{equation}
\nn   = \frac{E}{1-E} \sim   \frac{\Gamma(1/2)}{C} \frac{\sqrt{S}}{\ln S} .
\end{equation}
In addition, for the generating function of $\pi(S)$
\begin{align}
\Pi(z) = \frac{z\mathcal{F}'(z)}{1-z} \sim -\frac{C \ln(1-z)}{2(1-z)^{3/2}}.
\end{align}
Therefore,
\begin{align}
\pi(S) \sim \frac{C}{2\Gamma(3/2)}\sqrt{S} \ln S .
\end{align}
Then
\begin{align}
T(S) = \frac{ \pi(S)}{1-E(S)}+S \sim 2S.
\end{align}


\section{Asymptotic behavior for small $f$ for a random forager with random spread of food} \label{sec: app small f}
Let us analyze Eq. \eqref{eq: G random} in the limit of small density $f$,
\begin{equation}
\begin{gathered}
	\mathcal{F}(z) = G(\alpha(z)) = f \frac{\alpha}{1-\alpha +f\alpha } = \frac{f \alpha}{1-\alpha} \frac{1}{1+f \alpha /(1-\alpha)} 
	\\ 
	\sim \frac{f \alpha}{1-\alpha} - \left(\frac{f \alpha}{1-\alpha}\right)^2 \sim f \frac{ \alpha}{1-\alpha}.
\end{gathered}
\end{equation}
Consequently,
\begin{gather}
\mathcal{E}(z) = \frac{\mathcal{F}(z)}{1-z}\sim f \frac{ \alpha}{1-\alpha} \frac{1}{1-z} .
\end{gather}
Thus,
\begin{equation}
E \sim f.
\end{equation}
Hence,
\begin{equation}
\nn   = \frac{E}{1-E} \sim   f.
\end{equation}
For $\pi(S)$ we get
\begin{align}
\Pi(z) = \frac{z\mathcal{F}'(z)}{1-z} \sim f \frac{z}{1-z} \left( \frac{\alpha(z)}{1-\alpha(z)} \right)'.
\end{align}
Therefore,
\begin{equation}
\pi \sim f
\end{equation}
As a result,
\begin{gather}
\tau = \frac{\pi}{E} \sim Const. 
\end{gather}
For the lifetime
\begin{gather}
T - S = \tau \nn \sim f.
\end{gather}

\begin{figure}[h]
	\centering
	\includegraphics[width=1\linewidth]{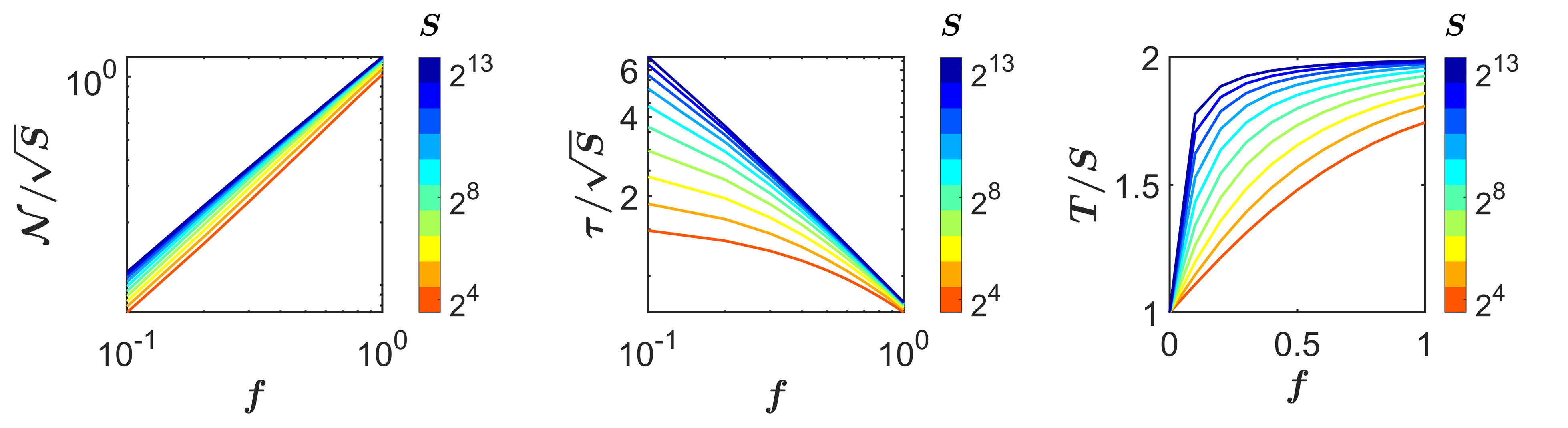}
	\caption{{\bf Random forager in one dimension with uniform random food distribution.} We can see the two limits of small $f$ and large ones. The quantity $\nn$ is linear with $f$ while $\tau$ approaches to a constant for small $f$ and then behaves like $1/f$, and $T$ is linear for small $f$ and then approaches to a constant.}
	\label{fig: random dist app}
\end{figure}

\FloatBarrier

\end{document}